\documentclass[letterpaper,english,aps,prb,twocolumn,floatfix,showpacs,amsfonts,amssymb,superscriptaddress]{revtex4}

\usepackage[T1]{fontenc}
\usepackage[latin1]{inputenc}
\usepackage{graphics}
\usepackage{amssymb}
\usepackage{amsmath}
\usepackage{epsfig}

\newcommand{\beq}{\begin{equation}}
\newcommand{\eeq}{\end{equation}}
\newcommand{\bea}{\begin{eqnarray}}
\newcommand{\eea}{\end{eqnarray}}

\newcommand{\rmi}{{\rm i}}

\newcommand{\cuoo}{CuO$_{2}$}
\newcommand{\lco}{La$_{2}$CuO$_{4}$}
\newcommand{\lasco}{La$_{2-x}$Sr$_{x}$CuO$_{4}$}
\newcommand{\lasnco}{La$_{2-x-y}$Nd$_{y}$Sr$_{x}$CuO$_{4}$}
\newcommand{\srcuocl}{Sr$_2$CuO$_2$Cl$_2$}

\begin{document}
\def\tende#1{\,\vtop{\ialign{##\crcr\rightarrowfill\crcr
\noalign{\kern-1pt\nointerlineskip} \hskip3.pt${\scriptstyle
#1}$\hskip3.pt\crcr}}\,}

\title{Magneto-elastic coupling between copper spin configurations and
  oxygen octahedra in {\lasco}. Untwinning, Raman (phonon) spectrum,
  and neutron response} 

\author{Marcello B. Silva Neto}

\affiliation{Institute for Theoretical Physics, University of
Utrecht, Leuvenlaan 4, 3584 CE Utrecht, The Netherlands.}

\begin{abstract}
We reinterpret neutron scattering and Raman spectroscopy experiments 
in {\lasco}, showing modulation of spin and lattice degrees of freedom 
both in the low-temperature orthorhombic and tetragonal phases, in
terms of a magneto-elastic coupling between noncollinear configurations 
for the copper spins and the oxygen octahedra. The magneto-elastic
coupling furthermore explains the recently discovered magnetic-field 
induced untwinning in {\lasco}, for $x<2\%$, and allows us to
understand why in {\lasnco} with $y=0.4$ and $x=1/8$, for example, 
when $(1,0)$ and/or $(0,1)$ static spin spirals are stabilized due to 
hole frustration, the atomic modulation in the low-temperature 
tetragonal phase of {\lasnco} has a periodicity of $4$ lattice 
spacings, half the magnetic one.
\end{abstract}

\maketitle

\section{Introduction}

{\lasco} is one of the best studied high
temperature superconductors and is a system where the phenomenon
of 1D charge segregation, or stripes formation, has been argued to
occur \cite{tranq}. In fact, inelastic neutron scattering
experiments within the superconducting phase of {\lasco},
$x>5.5\%$, have revealed that dynamical incommensurate (IC) spin
correlations coexist with superconductivity \cite{yamada}. In
addition, it has also been observed experimentally the existence
of IC "charge" peaks accompanying the IC magnetic order (even
though neutrons cannot directly probe charge distributions), with
twice the incommensurability of the magnetic one. This phenomenon
is most clearly observed when a low-temperature tetragonal (LTT)
phase is stabilized, for example by Nd doping, \cite{tranq} and it
was immediately interpreted as an evidence of the formation of
charge stripes that act as antiphase domain walls. \cite{zaanen}
In the same spirit, the observation of {\it diagonal} static IC
magnetic correlations at even lower doping, $2\%<x<5.5\%$, within
the low-temperature orthorhombic (LTO) spin-glass (SG) phase of
{\lasco}, \cite{NSSG} as well as recent evidence of diagonal
"charge" modulation from Raman spectroscopy, \cite{Tassini} have
also been used to support the picture of {\it diagonal} stripe
formation at lower doping.

In this article we propose an alternative interpretation for the
above described experimental results as originating from a
magneto-elastic coupling between the Cu$^{++}$ spins and the 
lattice in {\lasco}.\cite{Kastner} We start by explaining the 
recently discovered magnetic-field induced untwinning in {\lasco}.
\cite{Magnetoelastic} According to Ref. \onlinecite{Magnetoelastic}, 
the two inequivalent orthorhombic directions of {\lco} can be 
swaped at room temperature by the application of a strong in-plane 
magnetic field, $B\sim 14$ T. This remarkable effect was confirmed 
by X-ray measurements showing changes in the structure of {\lasco}, 
and the crystallographic rearrangement observed in an optical 
microscope.\cite{Magnetoelastic} This effect has been used since 
then to produce extremely high quality untwinned {\lasco} single 
crystals. We then argue that it is precisely this same magneto-elastic 
coupling that gives rise to a supermodulation of the oxygen 
octahedra when a noncollinear configuration for the Cu$^{++}$ 
magnetic moments is stabilized, as a consequence of the magnetic 
frustration introduced by hole-doping,\cite{SS} and we discuss 
its effects on the Raman and neutron responses, for the case of 
noncollinear magnetic structures both within the LTO and LTT 
phases of {\lasco}.

\section{The magneto-elastic coupling}

\begin{figure}[htb]
\begin{center}
\includegraphics[scale=0.26]{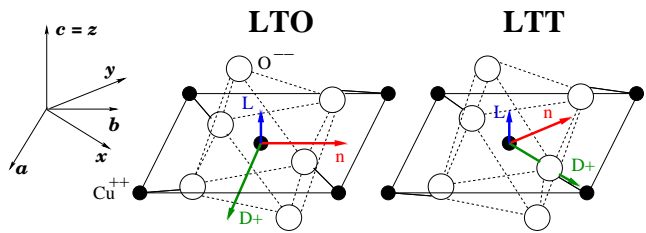}
\caption{(Color online) Left: LTO $(abc)$ and LTT $(xyz)$ coordinate
systems. Center and Right: Tilting of the octahedron in the LTO
(center) and LTT (right) phases. Uniform spin component ${\bf L}$ 
(blue arrow), staggered spin component ${\bf n}$ (red arrow), and
Dzyaloshinskii-Moriya vector ${\bf D}_+$ (green arrow).} 
\label{Fig-Tilting}
\end{center}
\end{figure}

In {\lco} each Cu$^{++}$ ion
is located at the center of an oxygen octahedron, see Fig.\
\ref{Fig-Tilting}. The crystal field produced by the O$^{--}$ ions
of the octahedron splits the energy levels of the Cu$^{++}$
orbitals in such a way that the unpaired electron is found at the
$d_{x^2-y^2}$ orbital. This is a planar orbital located at the
base of the octahedron (the common base of its two pyramids). The
spin-orbit (SO) interaction in {\lco} then introduces a large XY
anisotropy that favors a configuration for the Cu$^{++}$ spins
parallel to the plane of the $d_{x^2-y^2}$ orbital. In case there
is no tilting of the octahedra, this corresponds to confining the
spins to the {\cuoo} layers, as it happens for example in the
{\srcuocl} system. On the other hand, for {\lco} the tilting of
the oxygen octahedra in the LTO phase generates a
Dzyaloshinskii-Moriya interaction that determines an easy-axis for
the antiferromagnetic ordering and leads to the canting of the
spins out of the {\cuoo} planes, see Fig.\ \ref{Fig-Tilting}. This
can be better understood by using the notation introduced in Ref.\
\onlinecite{MLVC}. We first write the spin of Cu$^{++}$ in terms of its
staggered, ${\bf n}$, and uniform, ${\bf L}$, components, ${\bf
S}_i/S = e^{\rmi {\bf Q}\cdot {\bf x}_i}{\bf n}({\bf x}_i)+{\bf
L}({\bf x}_i)$, where, as usual, ${\bf Q}=(\pi,\pi)$ is the
antiferromagnetic ordering wave vector. Then we introduce a
thermodynamic Dzyaloshinskii-Moriya vector ${\bf D}_+$ pointing
along the axis of tilting of the oxygen octahedra, see Fig.\
\ref{Fig-Tilting}. 

The magnetic free-energy density for uniform
configurations of the staggered moments in a single layer of
{\lco} is\cite{MLVC}
\beq \small {\cal F}_M=\frac{1}{2gc} \left\{({\bf n}\cdot{\bf
D}_+)^2+\Gamma_{XY}{\;}n_c^2+({\bf n}\cdot{\bf B})^2-2{\bf
B}\cdot({\bf n}\times{\bf D}_+)\right\},\label{Free-Energy}\eeq
where $gc=8Ja^2$, $\Gamma_{XY}$ is the XY anisotropy energy, $J$
is the in-plane antiferromagnetic superexchange, and $a$ is the
lattice spacing. We now introduce a lattice free-energy density
\beq
{\cal F}_L=-\zeta{\;}{\bf D}_+\cdot \hat{\bf X},
\eeq
which favours ${\bf D}_+$ oriented along a certain direction 
$\hat{\bf X}$ in a fixed laboratory reference frame. The parameter
$\zeta$ introduced here controlls the softness of the oxygen octahedra
and in the case $\zeta\rightarrow\infty$ the octahedra becomes
completely rigid and fixed to the laboratory frame. 

Since the orientation of ${\bf D}_+$ defines the local orthorhombic  
coordinate system of the crystal, we conclude that, at zero magnetic 
field, ${\bf B}=0$, the equilibrium configuration of the system is
such that ${\bf D}_+\parallel a\parallel\hat{\bf X}$ with
\beq {\bf n}({\bf x}_i)\cdot{\bf D}_+=0, \quad \mbox{and} \quad
{\bf L}({\bf x}_i)=\frac{1}{2J}\left({\bf D}_+\times{\bf n}({\bf
x}_i)\right). 
\label{Eqs-Magnetoelastic} 
\eeq
Together with the large XY anisotropy, which confines ${\bf
n}({\bf x}_i)$ to the orthorhombic $ab$ basal-plane, the above two
conditions uniquely determine the easy-axis and canting of the
Cu$^{++}$ spins, with ${\bf n}\parallel b$ and ${\bf L}\parallel
c$, see Fig.\ \ref{Fig-Tilting}.

%
%

\section{Magnetic untwinning}

\begin{figure}[htb]
\begin{center}
\includegraphics[scale=0.35]{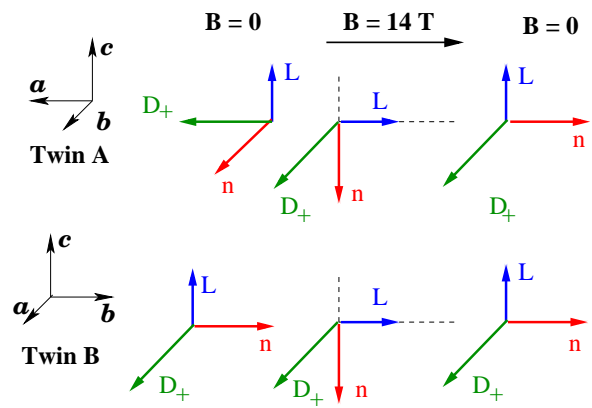}
\caption{(Color online) i) ${\bf B}=0$ (left): the two twins
have basis-vectors rotated by $\pi/2$ relatively; ii) 
${\bf B}=14$ T: the basis-vectors in twin A swap due to
the flop of staggered field ${\bf n}$ and orient 
identically to the ones of twin B, where the staggered ${\bf n}$ 
and uniform ${\bf L}$ vectors are rotated in the $bc$ plane 
(see discussion in the text); iii) ${\bf B}=0$ (right): the 
staggered ${\bf n}$ and uniform ${\bf L}$ vectors return to 
their equilibrium positions and the crystal remains untwined.}
\label{Fig-Swap}
\end{center}
\end{figure}

The physics of the magneto-elastic coupling in {\lasco} is
essentially given by Eqs.\ (\ref{Free-Energy}) and
(\ref{Eqs-Magnetoelastic}). These equations imply that the vectors
${\bf n}$, ${\bf L}$, and ${\bf D}_+$, form a triad of vectors
mutually orthogonal that can only move rigidly, see Fig.\
\ref{Fig-Tilting}. Let us now consider twins A and B such that the
orthorhombic basis-vectors in twin A are rotated by $\pi/2$
relative to the ones in twin B, $\hat{\bf X}_A\perp\hat{\bf X}_B$, 
see Fig.\ \ref{Fig-Swap}. When a strong in-plane magnetic field, 
$B>2gc\zeta$, is applied parallel to the $\hat{\bf X}_A$ orthorhombic 
direction of twin A, we find that the total free energy 
${\cal F}_M+{\cal F}_L$ is minimized, within such twin domain, 
for a new configuration of ${\bf n}$ and ${\bf D}_+$ such that 
${\bf n}\parallel c_A$ and ${\bf D}_+\parallel b_A$ (see 
Fig.\ \ref{Fig-Swap}, middle pannel). This is only possible if 
accompanied by a $\pi/2$ rotation of the entire octahedra in twin 
A (remember that ${\bf D}_+$ determines the axis of the tilting 
of the octahedra), which then causes the swap of the local 
orthorhombic directions. In twin B the effect of the same magnetic 
field is harmless, since it is longitudinal, 
${\bf B}\perp\hat{\bf X}_B$. In fact, a longitudinal 
magnetic field in {\lco} causes a continuous rotation of the two 
vectors, ${\bf n}$ and ${\bf L}$, within the $bc$ plane,
\cite{Marcello-Lara,Keimer} as a result of  the term 
$-2{\bf B}\cdot({\bf n}\times{\bf D}_+)$ in Eq.\ (\ref{Free-Energy}). 
Since in twin B the ${\bf D}_+$ vector remains perpendicular to 
the $bc$ plane, such rotation can take place without causing 
structural changes. When the field is removed, ${\bf n}$ and 
${\bf L}$ return to their original positions (as originally in 
twin B), but the orthorhombic axis in twin A remain swaped.

%
%

The important message from this magneto-elastic allowed untwining
in {\lasco} is that {\it a change in the spin configuration can 
have a strong effect on the equilibrium position of the tilting
axis of the oxygen octahedra}. What is most surprising, however, are 
the associated energy scales, especially being an effect originated 
from SO coupling. Besides the ability of causing the swap of the
crystallographic axis in a magnetic field (the sample actually
expanded about $1\%$ along the orthorhombic $b$ direction after
the untwinning\cite{Magnetoelastic}) magnetic susceptibility
experiments in {\lasco} by Lavrov {\it et al.}, \cite{Lavrov} at
much lower fields, have shown that the Cu$^{++}$ spins remain
confined to the orthorhombic $(bc)$ plane, for different doping
concentrations, $0<x<4\%$, and for a rather wide temperature
range, $0<T<400$ K. \cite{Lavrov} This is another consequence of
Eqs.\ (\ref{Eqs-Magnetoelastic}) that dictates that the ${\bf n}$
and ${\bf D}_+$ vectors should be perpendicular. \cite{MLVC}
In what follows we shall discuss the effects of the magneto-elastic 
coupling on the oxygen octahedra and its effect on the Raman and
neutron responses, for the case of noncollinear magnetic structures 
both within the LTO and LTT phases.

\section{Raman spectrum in the LTO phase}

In a recent publication,\cite{Helicoid} it was shown that the 
magnetic frustration introduced by hole-doping in the LTO phase 
of {\lasco} gives rise, above a certain critical doping, to a 
static helicoidal supermodulation where the ${\bf n}$ vector 
precesses (in space) around the $b$ easy-axis
\beq {\bf n}({\bf x}_i)=(\sigma_a\cos{({\bf Q}_S\cdot{\bf
x}_i)},\; n_b({\bf x}_i),\;\sigma_c\sin{({\bf Q}_S\cdot{\bf
x}_i)}), 
\label{Eqs-Helicoid} 
\eeq
with the wave vector for the spin incommensurability 
${\bf Q}_S\parallel b$, $\sigma_c\ll\sigma_a\ll 1$, and
$n_b^2=1-n_a^2-n_c^2$.\cite{Helicoid} Such helicoidal structure
was then shown to be consistent with the magnetic susceptibility 
experiments in {\lasco} for $x=2\%,3\%$ and $4\%$, since it 
guarantees that the total magnetization in a applied field 
remains confined to the $bc$ plane.\cite{Helicoid} 

\begin{figure}[htb]
\begin{center}
\includegraphics[scale=0.26]{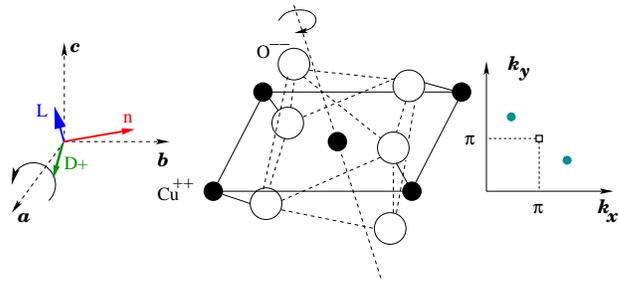}
\caption{(Color online) Precession of the triad (left)
and octahedron (center) around their equilibrium $(x=0)$
position. The precession takes place as one goes along the $b$
axis, such that the two wave vectors ${\bf Q}_S$ and ${\bf Q}_C$
are parallel to $b$. Right: neutron response in the LTO phase due to
modulation of ${\bf n}$. The neutron response is shifted from the 
N\'eel peak (unfilled square at $(\pi,\pi)$) to new diagonal
incommensurate positions (filled circles), due to the helicoidal
magnetic modulation described by Eq.\ \ref{Eqs-Helicoid}.}
\label{Fig-Precession}
\end{center}
\end{figure}

The physical picture behind the formation of such helicoidal 
structure can be easily understood as follows. Above a certain 
critical doping, that is essentially determined by the strength 
of the Dzyaloshinskii-Moriya gap, $\Delta_{DM}/J\sim 0.02$, the 
magnetic frustration introduced by the doped-holes favours a 
noncollinear configuration for the Cu$^{++}$ spins. Since these 
are coupled to the oxygen octahedra by the magneto-elastic 
coupling in Eq. (\ref{Eqs-Magnetoelastic}), any spin rotation 
should be accompanied by a rotation of the tilting axis of the
octahedra (${\bf D}_+$ is now ${\bf D}_+({\bf x}_i)$). However, 
in the LTO phase of {\lasco}, there are no $n$-fold axis of rotation 
with $n>2$, and thus the orthorhombicity frustrates any such 
rotation. For example, the in-plane spiraling of the ${\bf n}$ 
vector is forbidden because it would require a continuous 
rotation of the tilting axis of the octahedra around the $c$ 
axis and this is not allowed in the LTO phase. Thus, for Sr 
concentration well above the critical doping, deep inside the
SG phase, the resulting noncollinear spin structure in the LTO 
phase corresponds to, for example, a helicoidal modulation of 
the ${\bf n}$ vector around the $b$ easy-axis, as in 
Eq.\ (\ref{Eqs-Helicoid}), accompanied by a precession of the 
octahedron around its equilibrium $(x=0)$ position, see 
Fig.\ \ref{Fig-Precession}. Furthermore, such atomic modulation has 
wave vector for atomic incommensurability given by 
${\bf Q}_C\parallel b$. 

It is worth pointing out that, for even lower doping, 
$x\approx 0.01, 0.024$, it has been recently shown by 
Luscher {\it et al.}\cite{L-S} that the appropriate noncollinear
spin structure is such that the ${\bf n}$ field instead 
oscillates within the {\cuoo} plane around the $b$ easy-axis. 
We emphasize here that, once the magneto-elastic coupling in 
Eq. (\ref{Eqs-Magnetoelastic}) is taken into account, such 
oscillation also gives rise to a modulation of the octahedra, 
just as in the case of the helicoid.

%
%

\begin{figure}[htb]
\begin{center}
\includegraphics[scale=0.42]{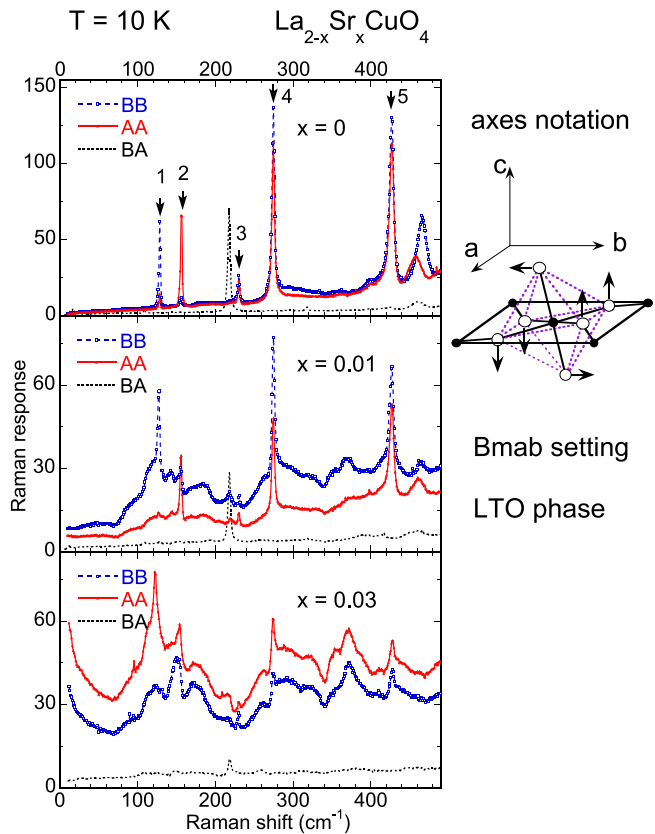}
\caption{(Color online) Raman response at $10$ K in the $(AA)$ 
(red-solid), $(BB)$ (blue-dashed), and $(BA)$ (black-dotted) 
polarization configurations, for $x=0,1\%$, and $3\%$ (from 
Ref.\ \onlinecite{Adrian}). For $x=0$ we clearly see sharp phonon 
peaks corresponding to the $5$ symmetric Raman active modes 
predicted by group theory, as well as one $B_{1g}$ phonon. 
For $x=1\%$ and $3\%$, however, a larger number of peaks 
is observed, signaling the enlargement of the atomic unit 
cell.} 
\label{Fig-Raman-LTO}
\end{center}
\end{figure}

The magnetic modulation with wave vector ${\bf
Q}_S\parallel b$, either helicoidal\cite{Helicoid} or 
oscillatory\cite{L-S}, gives rise to the diagonal IC 
peaks observed in the neutron response,\cite{NSSG} 
see Fig.\ \ref{Fig-Precession}. A supermodulation of the 
octahedron, in turn, would cause the enlargement of the 
atomic unit cell. This would result in the appearance of 
a plethora of extra vibrational modes that could be 
accessed via Raman spectroscopy.\cite{Adrian} As we 
shall now demonstrate, this is actually the case. On 
the top of Fig.\ \ref{Fig-Raman-LTO}, $x=0$, we observe $5$ 
sharp phonon peaks, corresponding to the symmetric vibrational 
modes predicted by group theory and accessible in the $(AA)$ 
(red-solid-curve) and $(BB)$ (blue-dashed-curve) Raman 
geometries. We also observe a sharp $B_{1g}$ phonon signal 
at $220$ cm$^{-1}$, accessible in the $(BA)$ (black-dotted-curve)
geometry. As doping concentration is increased, $x=1\%$ and 
$3\%$, the $5$ sharp phonon peaks are replaced by much more 
numerous and broader ones while the $B_{1g}$ phonon remains 
sharp, although with decreasing intensity, see 
Fig. \ref{Fig-Raman-LTO}. This is a consequence of 
the enlargement of the atomic unit cell associated
with the supermodulation of the octahedra with wave vector
${\bf Q}_C\parallel b$ (${\bf Q}_C$ is the wave vector for the
modulation of ${\bf D}_+({\bf x}_i)$), see Fig.\ \ref{Fig-Precession}.

%
%

In fact, recent Raman experiments in {\lasco} at $x=2\%$ showed
indications of anisotropic atomic modulation in the $B_{2g}$ Raman
geometry. \cite{Tassini} These results are consistent with the
above physical picture of the octahedra supermodulation. The
atomic modulation resulting from the precession of the octahedra
has wave vector ${\bf Q}_C\parallel b$. In order to probe such
modulation, we must choose a Raman geometry such that the wave
vector ${\bf Q}_C$ has a finite projection on the axis of the
electric field of both incoming $e_{in}$ and outgoing $e_{out}$
light. Thus, the above octahedra supermodulation should become
accessible exactly in the $B_{2g}$ geometry, where $e_{in}$ and
$e_{out}$ are oriented along the Cu-Cu bonds, with respect to
which ${\bf Q}_C$ is diagonal. \cite{Tassini} Furthermore, it
should not be accessible at all in the $(BA)$ geometry
(black-dotted-curve) where $e_{in}$ and $e_{out}$ are oriented 
along the orthorhombic basis-vectors, and this results in a 
rather featureless $B_{1g}$ response (except for the phonon at 
$220$ cm$^{-1}$), see Fig.\ \ref{Fig-Raman-LTO}. Although the 
results by Tassini {\it et al.} \cite{Tassini} were interpreted 
as a signature of diagonal stripe formation, we believe that 
the octahedra supermodulation provides a more reasonable 
explanation for the Raman response reported in 
Fig.\ \ref{Fig-Raman-LTO} and in Ref.\ \onlinecite{Tassini}, 
especially for the $x=1\%,3\%$ samples where stripes can be safely 
ruled out.

\section{Neutron response in the LTT phase}

\begin{figure}[htb]
\begin{center}
\includegraphics[scale=0.4]{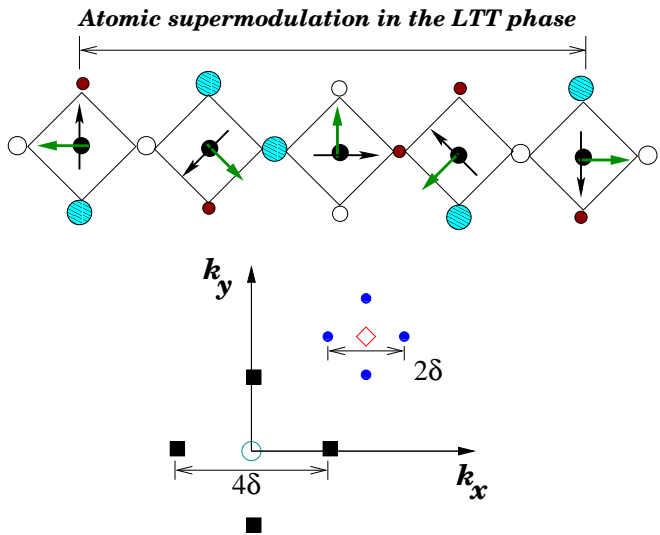}
\caption{(Color online) Top: Magneto-elastic coupled spiraling of the
Cu$^{++}$ spins (black arrows, V-shaped tip) and ${\bf D}_+$ 
(green arrow, triangular tip). Blue (larger and hatched) oxygen 
atoms are canted up the plane, red (smaller and filled) ones are 
canted down the plane, and open ones are confined to the basal 
plane. The octahedra on the far left and far right, exactly at 
half the magnetic periodicity, are equivalent (see discussion in 
the text). Bottom: Since the atomic periodicity is half the magnetic 
one, the neutron response in reciprocal space is such that the atomic
incommensurability (large filled squares) is twice the magnetic 
one (smaller filled circles).} 
\label{Fig-Rotation-Octahedra}
\end{center}
\end{figure}

When a tetragonal phase is stabilized in {\lasco}, by Nd doping 
for example, incommensurate "charge" peaks are observed in neutron 
scattering, whose incommensurability is twice the magnetic one, 
see Fig.\ \ref{Fig-Rotation-Octahedra}. Even though neutrons cannot 
directly couple to charge excitations, this result has been 
interpreted, for over 10 years, as a sign of charge stripe formation 
acting as anti-phase domain walls.

Let us now demonstrate how the magneto-elastic Eqs.\
(\ref{Eqs-Magnetoelastic}) can be used to explain the neutron
scattering results within the LTT phase. In the LTT phase, the
${\bf D}_+$ vector points along one of the basal tetragonal
directions, say $x$, such that two oxygen atoms remain confined to
the tetragonal basal plane while the other two are tilted, one
above and one below the {\cuoo} layers, see Fig.\
\ref{Fig-Tilting}. From the magneto-elastic coupling 
equations Eq.\ (\ref{Eqs-Magnetoelastic})
we then find ${\bf n}\parallel y$ and ${\bf L}\parallel z$.
There is one major difference, however, regarding possibilities for the
rotation of the tilting axis of the octahedron between the 
LTO and LTT phases when a noncollinear spin structure is favoured. 
In the LTT phase, the $z$ axis is in fact a $n$-fold axis with $n>2$, 
and a in-plane rotation of the staggered moments is now not only 
allowed but energetically favourable (as discussed earlier, the XY 
anisotropy favours ${\bf n}({\bf x}_i)$ to be parallel to the
planes). The above argument is consistent with the calculations by 
Sushkov and Kotov, \cite{SK} using realistic parameters for the 
$t-t^\prime-t^{\prime\prime}-J$ model, which showed that the magnetic 
ground state, in the LTT phase, is composed by $(1,0)$ and/or $(0,1)$ 
spirals
\beq {\bf n}({\bf x}_i)=(\cos{({\bf Q}_S\cdot{\bf x}_i)},\;
\sin{({\bf Q}_S\cdot{\bf x}_i)},\; 0), \label{Eqs-Spiral} \eeq
with $Q_S=\delta\sim x$. The magneto-elastic coupled spiraling of the
Cu$^{++}$ spins and tilting axis of the octahedra, ${\bf D}_+$,
produces the pattern depicted in Fig.\ \ref{Fig-Rotation-Octahedra}.

%
%

Now, by analyzing Fig. \ref{Fig-Rotation-Octahedra} we can finally
understand why the atomic incommensurability is twice the magnetic
one, and why for $x=1/8$ the atomic periodicity is of $4$ lattice
spacings. Since $Q_S=\delta\sim x$, for $x=1/8$ the spin
periodicity is of $8$ lattice spacings, obtained after $8$
rotations of $\pi/4$. \cite{SK} However, after $4$ rotations of
$\pi/4$, when the spins are anti-parallel, the configuration of
the octahedron is equivalent to the initial one, see Fig.\
\ref{Fig-Rotation-Octahedra}. Although this might not be apparent
on a first inspection due to the difference in the tilting of the
oxygen atoms, we should not forget that in a body-centered
tetragonal unit cell, such as the one of {\lasco}, corner and
central Cu$^{++}$ ions are equivalent. Thus, any symmetry
operation can be defined together with a half translation of the
lattice vectors. This brings the octahedron on the far right of
Fig.\ \ref{Fig-Rotation-Octahedra} to the one in the far left, and
so they are indeed equivalent. Notice now that the modulation of
the atoms in the octahedra is half the magnetic one, and so the
neutron response in the reciprocal space should be such that the
atomic incommensurability is twice the magnetic one, \cite{tranq}
see Fig.\ \ref{Fig-Rotation-Octahedra}.

Finally, the magneto-elastic coupling might also gives us a hint 
towards the understanding of some unexpected features on the phonon 
spectrum of the {\lasnco} system, at $x=1/8$ and $y=0.4$, recently 
reported by the Raman experiments of Gozar {\it et al.} in 
Ref.\ \onlinecite{Adrian}. It has been observed that the lowest 
energy of the $A_g$ phonon modes, there denoted $A$ mode, has a 
rather broad Raman spectral line. Since this is the phonon mode 
related to the spatial distribution of the tilting angle of the 
oxygen octahedra, the Raman data seems to indicate a locally
fluctuating octahedra tilt distribution in {\lasnco},\cite{Adrian} 
that can be understood in terms of short-range magneto-elastic 
coupled spin-tilt-axis fluctuations.

\section{Conclusions} 

We have shown that neutron scattering and
Raman spectroscopy experiments, either in the LTO or LTT phases of
{\lasco}, as well as the process of untwining in a strong in-plane
magnetic field, can be understood in terms of the unusually strong
magneto-elastic coupling between the noncollinear configuration for
the Cu$^{++}$ spin and the oxygen octahedra. 
Although the analysis here perfomed can go no further than discussing
static phenomena of charge and spin ordering, it is very much possible 
that the distribution of oxygen doped holes in {\lasco} will follow
the anisotropic 1D octahedra modulation proposed above, thus presenting 
us with an interesting possibility for the origin of the anisotropic 
1D electronic (charge) dynamics in cuprates. This interesting
possibility is presently under investigation.\cite{MBSN-Next} 

\section{Acknowledgements} 

The author acknowledges invaluable discussions with Y.~Ando,
L.~Benfatto, A.~H.~Castro~Neto, B.~Keimer, A.~Gozar, 
C.~Morais~Smith, and O.~Sushkov. A special thanks goes to 
A.~Gozar for kindly producing Fig.\ \ref{Fig-Raman-LTO}, 
containing the data published in Ref.\ \onlinecite{Adrian}.


\end{document}